\documentclass[12pt]{article}
\def\ee{\end{equation}}
\def\be{\begin{equation}}
\def\ba{\begin{eqnarray}}
\def\ea{\end{eqnarray}}
\def\ll{\label}
\title{Composable entropy and deviation from macroscopic equilibrium}
\author{Ramandeep S. Johal
\footnote{e-mail: rjohal@jla.vsnl.net.in} \\
643-L, Model Town, Jalandhar-144003, India.}
\begin{document}
\maketitle
\abstract{We formulate, under general conditions, the problem of maximisation of the total entropy of the system, assumed to be in a composable form, for fixed total value of the constrained quantity. We derive the general form of
the composability function and also point out the criterion which leads to a violation
of the zeroth law of thermodynamics.}
\newpage
\baselineskip 24pt
%
%
Macroscopic thermodynamics is based on an entropy function, which
is additive with respect to independent subsystems \cite{Callen1985}. The theory of
statistical mechanics, which provides microscopic foundations
for thermodynamics, also naturally treats entropy as additive,
and predicts exponential distributions as the equilibrium distributions.
Recently, various generalized entropic functionals \cite{Tsallis1988,Lenzirenyi,
Gorban2003}, which may or may not be 
additive with respect to independent subsystems, have been proposed 
in order to generalize the statistical mechanical formalism. This is in part 
motivated by the observation that in nature very often, non-exponential,
non-gaussian and power law distributions better describe the
statistical properties of complex phenomena.

Along with these studies, it is also of high current interest to 
understand as to which general forms of entropy are 
consistent with the thermodynamical framework. 
Recently, the notion of composability has been found useful to
understand these issues \cite{Hotta1999,Abepre01,Wangjpa2002}.
A generalized entropy, when in a composable form,
implies that  the total entropy
of the composite system made up of say, two subsystems $A$ and $B$, can be
written as 
\be
S(A,B) = f(S(A),S(B)),
\ll{compose}
\ee
where $f$ is a certain bivariate function, such that the function
itself and all its derivatives are continuous. Moreover, $f$ is symmetric in its
arguments: 
\be
f(S(A),S(B)) =  f(S(B),S(A)).
\ll{absymmetry}
\ee
The function $f$ satisfies the following natural properties:
\be
f(0,S(B)) = S(B),  \quad f(S(A),0) = S(A),
\label{f0}
\ee
and 
\be
f(0,0) = 0.
\label{f00}
\ee
Some known forms of the function $f$, which have been studied in literature are:
\be
S(A,B) = S(A) + S(B),
\ll{sadditive}
\ee
which is  obeyed, for instance, by entropic forms proposed by Renyi  
and Shannon.
On the other hand, certain non-additive 
entropic functionals are also in vogue these days. Tsallis 
entropy \cite{Tsallis1988}, for instance, obeys 
\be
S(A,B) = S(A) + S(B) + \omega S(A) S(B),
\ll{stsallis}
\ee
where the real parameter $\omega$ represents the degree of nonadditivity.
Tsallis entropy is also obviously in a composable form.
Many papers have been devoted recently \cite{Tsallishttp},
to study how laws and various
thermodynamic relations are generalized or left invariant when 
Tsallis entropy is employed.
The non-exponential distributions predicted by employing the maximum entropy
variational principle, have been applied to the stationary states of
certain nonextensive systems.

The concept of entropic 
composability puts a stringent constraint on the form
of total entropy. An important question is what kind of
composability functions $f$ make the entropy consistent with
laws of thermodynamics. It is evident that composability does
not by itself guarantee this consistency. 
A step was made in this direction in Ref. \cite{Abepre01}
by showing that the Tsallis kind of nonadditivity is the simplest
example of composability, that may be compatible with the
existence of thermodynamic equilibrium, specifically, with the zeroth law
of thermodynamics. In this paper, we revisit this approach
and generalise it further in order to analyse as to when
a composable entropic form may not follow 
 the zeroth law of thermodynamics.

First, let us review the approach adopted in Ref. \cite{Abepre01}.
 To be able to formulate thermodynamics,
the simplest condition one can put is to maximize the
total entropy (\ref{compose}),  subject to some additive constraints.

Let the only constraint be expressed as 
\be
E(A,B) = E(A) + E(B),
\label{constraint}
\ee                                                    
where the quantity $E$ may represent the internal energy. 
By making the variations of the total
entropy, $dS(A,B)$, and of the total value of the constraint quantity, $dE(A,B)$,
vanish, we get
\be
{\partial {S}(A, B) \over
\partial {S}(A) } {\partial {S}(A)
\over \partial E(A)} =
{\partial {S}(A, B) \over
\partial {S}(B) } {\partial {S}(B)
\over \partial E(B)}.
\label{var0}
\ee                                               
Now if the entropy function is simply additive (Eq. (\ref{sadditive})),
then the first partial derivative on either side of the above equation
is identically equal to unity. In this standard case, after defining a
quantity called temperature $T= (\partial S/ \partial E)^{-1}$,
this equation yields that the temperatures
of the two subsystems are equal. This condition defines the state of thermal
equilibrium between the subsystems, in macroscopic thermodynamics \cite{Callen1985}.

To obtain such an equality condition for the more general case of a 
composable entropy, an additional condition was assumed \cite{Abepre01}:  the
relation (\ref{var0}) should yield a {\it separable} system of equations.
To accomplish this, the following factorizability condition is a natural 
choice \cite{Abepre01}:
\ba
{\partial {S}(A, B) \over \partial {S}(A) }
= g[ S(A)] h[S(B)], \ll{separableab} \\
{\partial {S}(A, B) \over \partial {S}(B) }
= g[ S(B)] h[S(A)],
\ll{separableba} 
\ea
where $g$ and $h$ are some functions. In particular, $h$ has to be
a differentiable function. 
Clearly, using the above equations in (\ref{var0}) and rearranging,
we can obtain separation of variables of the form $F(A)=F(B)$. The
hope is that we can then identify temperature-like quantity for
the subsystems and thus arrive at a generalized version of the
zeroth law. We reserve further remarks about this approach till
the discussion in the end of the paper. 

Proceeding further,  for any subsystem, say $A$, one can
then show that
\be
g[S(A)] = \frac{1}{\omega} \frac{ dh[S(A)]}{d S(A)}.
\label{gh}
\ee
Now Tsallis type nonadditivity (Eq. (\ref{stsallis}))
is consistent with this framework, if we identify
\be
h[S(A)] = 1 + \omega S(A),
\label{sa}
\ee
and similarly for system $B$. This yields $g[S(A)] = g[S(B)] = 1$.
Thus Eqs. (\ref{separableab}) and (\ref{separableba}) simplify to
\ba
{\partial {S}(A, B) \over \partial {S}(A) }
=  1+ \omega S(B), \ll{separableab2} \\
{\partial {S}(A, B) \over \partial {S}(B) }
= 1 + \omega  S(A).
\ll{separableba2} 
\ea
It may be important to emphasize again  that 
in standard thermodynamics, we consider additive nature of both total
entropy and the constraints. The  only further
assumption is that the total entropy of the composite system is maximized,
under a fixed total value of the constrained quantity ($E$).
Separability in the form of $F(A) = F(B)$, is achieved automatically there. 
The motivation of Ref. \cite{Abepre01} is to infer the form
of the composability function, 
by imposing further conditions like factorizability as in 
Eqs.~(\ref{separableab}) and (\ref{separableba}).
In our opinion, these conditions must restrict the possible 
forms of the composability function, that may be consistent with
the maximum  of the total entropy.

In this paper, following the spirit of standard 
thermodynamics, we study  the consequences of
just {\it one assumption: maximization of the entropy of the total system,
under a fixed total value of the constrained quantity}.
In other words, we do not start with the simple factorized forms
(\ref{separableab}) and (\ref{separableba}), but assume more general 
non-factorisable forms. As will become clear, we do however, 
make use of a separability criterion at a later stage. 
Our analysis not only incorporates Tsallis type of nonadditivity as a 
special case, but also naturally leads to further classes 
of the composability function, though they may not lead to
formulation of a zeroth law (or equality of temperatures for subsystems).
Thus we also arrive at a criterion as to when this violation
of the zeroth law may be expected, for a composable entropic form. 

A simple choice violating  the separability condition 
for (\ref{var0}) may be as follows:
\ba
{\partial {S}(A, B) \over \partial {S}(A) }
= l[ S(A)] m[S(B)] + n[S(B)], \ll{nonseparableab} \\
{\partial {S}(A, B) \over \partial {S}(B) }
= l[ S(B)] m[S(A)] + n[S(A)].
\ll{nonseparableba}
\ea
Here $l$, $m$ and $n$ are arbitrary continuous functions.
Particularly, $n$ may have continuous derivatives upto an 
arbitrary order. Using the fact that 
\be
{\partial^2 {S}(A, B) \over \partial {S}(B)\partial {S}(A) }
=
{\partial^2 {S}(A, B) \over \partial {S}(A)\partial {S}(B) },
\ll{abderivative}
\ee
we get
\be
l[S(A)] \frac{d m[S(B)]}{d S(B)} +  \frac{d n[S(B)]}{d S(B)}
=
l[S(B)] \frac{d m[S(A)]}{d S(A)} +  \frac{d n[S(A)]}{d S(A)}.
\ll{solvedd}
\ee
This is a relation between three unknown functions $l$, $m$ and $n$.
To analyse further, we have to introduce the following simplification.
Let $m$ be a linear function of its argument
\be
m[x] = a+b x.
\ll{defh}
\ee
A convenient choice is $a=0$, if we demand that $f(0,0) = S(A,B) =0$. 
Here, $b$ is another constant.
Then (\ref{solvedd}) is simplified to
\be
b\; l[S(A)]  +  \frac{d n[S(B)]}{d S(B)}=
b\; l[S(B)] +  \frac{d n[S(A)]}{d S(A)}.
\ll{simplegi}
\ee
This leads to a separation of variables and  can be written as
\be
b\; l[S(A)] -  \frac{d n[S(A)]}{d S(A)}=
b\; l[S(B)] -  \frac{d n[S(B)]}{d S(B)} = \omega,
\ll{separategi}
\ee
where $\omega$ is a constant of separation. 

Then using (\ref{defh})
and (\ref{separategi}), the conditions  (\ref{nonseparableab})
and (\ref{nonseparableba}) can be expressed as
\ba
{\partial {S}(A, B) \over \partial {S}(A) }
= \left( \frac{d n[S(A)]}{d S(A)} + \omega \right) S(B)  + n[S(B)],
\ll{finalconditionsab} \\
{\partial {S}(A, B) \over \partial {S}(B) }
= \left( \frac{d n[S(B)]}{d S(B)} + \omega \right) S(A) + n[S(A)],
\ll{finalconditionsba}
\ea
either of which may be integrated to give the form of 
composability function:
\be
f(S(A),S(B)) = S(A,B) = n[S(B)] S(A) + n[S(A)] S(B) +  \omega S(A) S(B).
\ll{solutionsab}
\ee
This function satisfies the symmetry condition, Eq.~(\ref{absymmetry}), 
as required. As special cases of this form of composability function,
let $n$ be a constant equal to unity. Then we obtain Tsallis
type nonadditivity of degree $\omega$. Alternately, we can set $\omega = 0$ and
discuss special cases of the following composable function
\be
f(S(A),S(B)) = S(A,B) = n[S(B)] S(A) + n[S(A)] S(B).
\ll{solutionsab2}
\ee
First note that, using the condition (\ref{f0}),
we obtain $n[0] =1$. Now

i) if $n$ is a constant, say equal to unity, we obtain the additivity
   of entropy. 

ii) if $n$ is a linear function, $n[x] = 1 + \lambda x$, we obtain
    Tsallis type of nonadditivity with degree $\lambda$.
    
iii) if $n$ is a nonlinear function of its argument, then
     separation of variables in the form $F(A)=F(B)$ in Eq. (\ref{var0}),
     cannot be achieved and thus we cannot arrive at the notion of
     equal temperatures for the two subsystems.
We illustrate this with an example. 
A composable and nonadditive entropic form was recently 
studied in Ref. \cite{Kaniadakispre2002}. This form was motivated
in the context of special relativity. It satisfies
\be
S(A,B) = S(A)\sqrt{1 + \kappa^2 S(B)^2} +  S(B)\sqrt{1 + \kappa^2 S(A)^2},
\ll{skania}
\ee
which goes to the additive form, when the real parameter $\kappa \to 0$.
However, the above form  is a particular instance of
(\ref{solutionsab2}) with  $n[x] = \sqrt{1 + \kappa^2 x^2}$.
The function $n$ satisfies the condition (iii) and  
clearly  therefore, the notion of equal temperatures cannot be
formulated with this function, upon maximisation of the total entropy.

Summarising, we have revisited the problem of maximisation of
a composable entropy under the fixed value of a constrained quantity
(total energy) \cite{Abepre01}. We have approached the problem 
from more general considerations and have not imposed the factorisation
condition as discussed in \cite{Abepre01}. The factorisation assumption
was crucial there in order to arrive at a generalised version of the
zeroth law when using composable entropic forms.
However, this approach has also been analysed further in many
papers, where a mapping to an additive entropy is shown \cite{
Johalpla03,Johalpp0}. In other
words, the separation of variables in the form $F(A)=F(B)$, may be 
interpreted as equivalent to the maximisation of an additive entropy
under fixed value of additive constraints, where then $F$ is identified
as the intensive variable. 
On the other hand, zeroth law is essentially valid for {\it macroscopic}
thermodynamics. It is known that for finite systems, the intensive
variables, such as temperature are not equal over the subsystems \cite{Hill1994,
Prosper93}.
In view of the growing
interest in finite systems \cite{Hill2001a,Hill2001b,Evans2nd}, 
it is of importance to investigate if
generalised entropies can  describe effects of finiteness \cite{Gorban2003}. 
The main motivation  of present work is to formulate and point out the
conditions which lead to violation of the zeroth law for a composable
entropy. Seen in this context, the present analysis may give
further clues about the relevance of composable entropies
in the description of finite systems.

%

\end{document}